\documentclass[twocolumns]{aa} % for the letters 
\usepackage{graphicx}
\usepackage{txfonts}
\usepackage[round]{natbib}
\begin{document}
   \title{Challenging shock models with SOFIA OH observations\\in the high-mass star-forming region Cepheus A}

   \author{A. Gusdorf
          \inst{1}
          \and
          R. G\"usten\inst{2}
          \and
          K. M. Menten\inst{2}
          \and
          D. R. Flower\inst{3}
          \and 
          G. Pineau des For\^ets\inst{4}
           \and
          C. Codella\inst{5}
          \and
          T. Csengeri\inst{2}
          \and
          \\A. I. G\'omez-Ruiz\inst{2}
          \and 
          S. Heyminck\inst{2}
          \and
          K. Jacobs\inst{6}
          \and
          L. E. Kristensen\inst{7}
	\and
          S. Leurini\inst{2}
          \and 
          M. A. Requena-Torres\inst{2}
           \and
          \\S. F. Wampfler\inst{8}
          \and
          H. Wiesemeyer\inst{2}
          \and
          F. Wyrowski\inst{2}
           }

  	   \institute{LERMA, Observatoire de Paris, PSL Research University, CNRS, UMR 8112, F-75014, Paris, France; Sorbonne Universit\'es, UPMC Univ. Paris 6, UMR 8112, LERMA, F-75005, Paris, France; \email{antoine.gusdorf@lra.ens.fr}
             \and
             Max Planck Institut f\"ur Radioastronomie, Auf dem H\"ugel, 69, 53121 Bonn, Germany
             \and
             Physics Department, The University, Durham DH1 3LE, UK
             \and
             IAS, UMR 8617 du CNRS, B\^atiment 121, universit\'e de Paris Sud, F-91405, Orsay, France
             \and
             INAF, Osservatorio Astrofisico di Arcetri, Largo Enrico Fermi 5, I-50125 Firenze, Italy   
             \and
             KOSMA, I. Physikalisches Institut, Universit\"at zu K\"oln, Z\"ulpicher Str. 77, 50937 K\"oln, Germany
             \and
             Harvard-Smithsonian Center for Astrophysics, 60 Garden Street, Cambridge, MA, 02138, USA   
             \and
             Centre for Star and Planet Formation, Niels Bohr Institute and Natural History Museum of Denmark, University of Copenhagen, \O ster Voldgade 5-7, DK-
1350, K\o benhavn K, Denmark}

%   \date{Received September 15, 1996; accepted March 16, 1997}

% \abstract{}{}{}{}{} 
% 5 {} token are mandatory
 
  \abstract
  % context heading (optional)
  % {} leave it empty if necessary  
   {OH is a key molecule in H$_2$O chemistry, a valuable tool for probing physical conditions, and an important contributor to the cooling of shock regions around high-mass protostars. OH participates in the re-distribution of energy from the protostar towards the surrounding Interstellar Medium.}
  % aims heading (mandatory)
   {Our aim is to assess the origin of the OH emission from the Cepheus A massive star-forming region and to constrain the physical conditions prevailing in the emitting gas. We thus want to probe the processes at work during the formation of massive stars.}
  % methods heading (mandatory)
   {We present spectrally resolved observations of OH towards the protostellar outflows region of Cepheus A with the GREAT spectrometer onboard the Stratospheric Observatory for Infrared Astronomy (SOFIA) telescope.  Three triplets were observed at 1834.7~GHz, 1837.8~GHz, and 2514.3~GHz (163.4~$\mu$m, 163.1~$\mu$m between the $^2\Pi_{1/2}$ $J = 3/2$ and $J = 1/2$ states, and 119.2~$\mu$m, a ground transition between the $^2\Pi_{3/2}$ $J = 5/2$ and $J = 3/2$ states), at angular resolutions of 16\farcs3, 16\farcs3, and 11\farcs9, respectively. We also present the CO (16--15) spectrum at the same position. We compared the integrated intensities in the redshifted wings to the results of shock models.}
  % results heading (mandatory)
   {The two OH triplets near 163~$\mu$m are detected in emission, but with blending hyperfine structure unresolved. Their profiles and that of CO (16--15) can be fitted by a combination of two or three Gaussians. The observed 119.2~$\mu$m triplet is seen in absorption, since its blending hyperfine structure is unresolved, but with three line-of-sight components and a blueshifted emission wing consistent with that of the other lines. The OH line wings are similar to those of CO, suggesting that they emanate from the same shocked structure.}
  % conclusions heading (optional), leave it empty if necessary 
   {Under this common origin assumption, the observations fall within the model predictions and within the range of use of our model only if we consider that four shock structures are caught in our beam. Overall, our comparisons suggest that all the observations might be consistently fitted by a J-type shock model with a high pre-shock density ($n_{\rm H} > 10^5$~cm$^{-3}$), a high shock velocity ($\varv_{\rm s} \gtrsim 25$~km~s$^{-1}$), and with a filling factor of the order of unity. Such a high pre-shock density is generally found in shocks associated to high-mass protostars, contrary to low-mass ones.}%Such shocks are dissociative, and might generate a radiative precursor in the pre-shock medium. They also require a detailed treatment of grain--grain interactions. The influence of a possible protostellar radiation field is not investigated.


  \authorrunning{A. Gusdorf et al.}
  \titlerunning{OH emission in the high-mass star forming region Cepheus A}
   \maketitle
%
%________________________________________________________________

\section{Introduction}
\label{sec:intro}

\begin{figure*}
\centering
\includegraphics[width=0.85\textwidth]{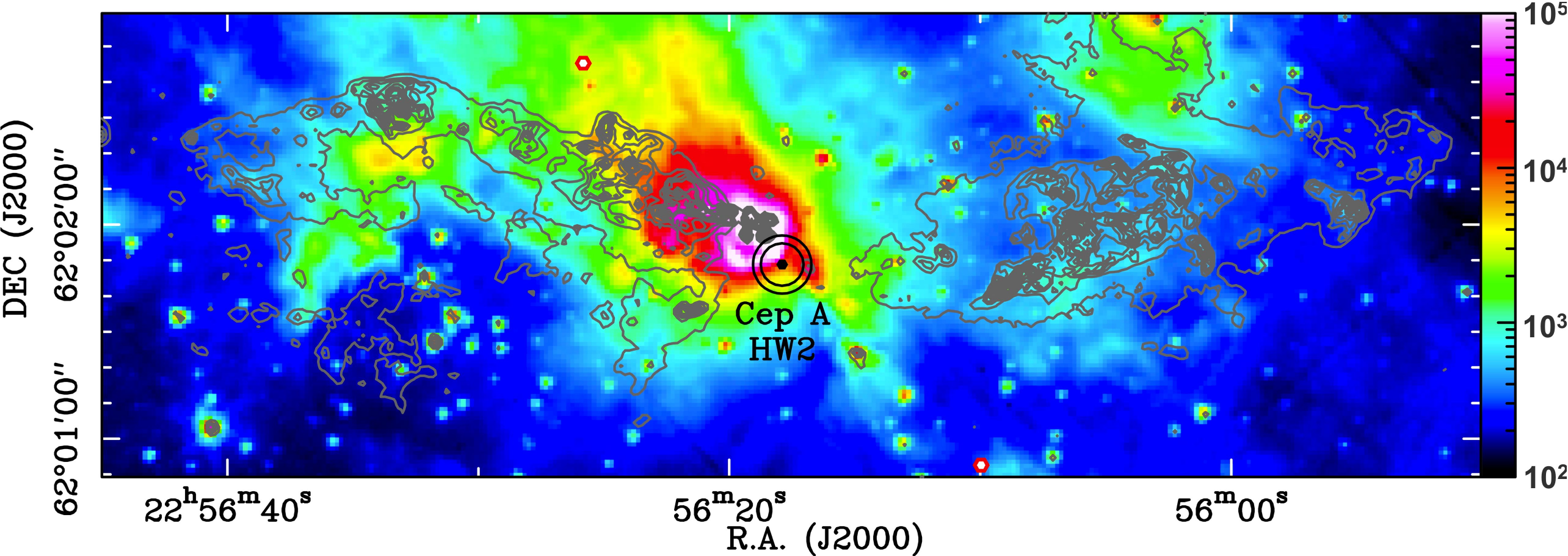}
\caption{Cepheus A region as seen in channel 1 of \textit{Spitzer}/IRAC at 3.6 $\mu$m (background colours; retrieved from the \textit{Spitzer} archive; wedge units are MJy/sr), and in H$_2$ at 2.12 $\mu$m (in dark grey contours, from \citealt{Cunningham09}). The observed position `Cep A HW2' is marked by a black hexagon, surrounded by the GREAT beam sizes in CO (16--15) (roughly the same as in OH 1835~GHz and 1838~GHz) and OH 2514~GHz (respectively 16\farcs3 and 11\farcs9). Multiple OH maser spots have been found in the vicinity of this source by \citet{Cohen84} and \citet{Bartkiewicz05}. The off positions are also indicated by white and red hexagons in the north-eastern and south-western directions with respect to the observed position.}
\label{figure1}
\end{figure*}

 Observations over the past few decades have shown that, in the early stages of star formation, the process of mass accretion is almost always associated with mass ejection in the form of collimated jets.  The jets impact on the parent cloud, driving a shock front through the collapsing interstellar gas. Large cavities, called bipolar outflows, are carved in the ambient medium, which is accelerated, compressed and heated by the shock wave. This paradigm was proposed a few decades ago by \citet{Snell80}, in connection with the formation of low-mass stars, and has been regularly verified in such environments (see, for example, \citealt{Arce07, Frank14} for reviews), including recent high-angular-resolution observations by ALMA (e.g. \citealt{Codella14}). However, establishing its applicability to the formation of massive stars remains a challenge for observers and modellers (see \citealt{Tan14} for a review of massive star formation). A central question is whether the shocks that are generated by massive protostars have similar physical and chemical properties to those driven by young stellar objects (YSOs) of lower mass. Studying the molecular emission from star-forming regions (SFRs) is a way to progress on this question. 
 
 In non-dissociative shock waves, the kinetic temperature of the gas can rise to a few thousand degrees, at which point the energy barriers to numerous chemical reactions can be overcome. Other processes affect the dust grains, resulting in a significant alteration of the abundances of certain species \citep{Bachiller01, Flower03}. Among the shock-tracing molecules, water is particularly important: it is a carrier of oxygen, a relatively abundant element that modifies the gas-phase or grain-surface chemistry of many other species, and is an important coolant of the gas (see, for example, \citealt{Vandishoeck11, Vandishoeck13, Vandishoeck14} for reviews of these aspects). If its chemistry were properly understood, H$_2$O would be a most appropriate molecule to comparatively study the formation of stars of various masses. However, this is not the case, and, with a view to understanding the abundance of H$_2$O and to characterizing the nature of the shock waves generated during the star formation process, the hydroxyl radical has emerged as a key species (see \citealt{Wampfler13}, hereafter W13, for an overview of previous studies of OH). OH is chemically linked to H$_2$O through the OH + H$_2 \Longleftrightarrow$ H$_2$O + H reactions. The formation of H$_2$O from OH is expected to be efficient in jets and outflows through both high-temperature gas-phase and grain-surface chemistry. Below around 250~K, \lq standard' gas-phase chemistry applies, in which H$_2$O is formed and destroyed principally through ion-molecule reactions. 

OH is a product of H$_2$O photodissociation, and some observational studies have linked a lack of H$_2$O directly to an enhanced abundance of OH. This photodissociation may be driven by the radiation field of the protostar, in low-mass YSOs (as mentioned in \citealt{Wampfler10, Karska13, Karska142}) and high-mass YSOs (as evoked in \citealt{Karska141} and \citealt{Wampfler11}, hereafter W11). It might also be driven by the radiation emitted by the shock wave itself in low-mass YSOs (as mentioned in \citealt{Wampfler10, Karska142}). For instance, in the bipolar outflow system HH~211, driven by a low-mass YSO, \citet{Tappe08,Tappe12} have attributed the detection of superthermal OH emission (i.e. resulting from the population of rotational levels up to at least ~28 200 K despite the high $A$ coefficeint values) to the photodissociation of H$_2$O by the UV radiation generated in the terminal shock. The photodissociation may also be triggered by an external source of radiation, such as an intermediate-mass YSO irradiating lower mass SFRs (\citealt{Lindberg14}). Alternatively, OH has been proposed as a tracer of dissociative shocks \citep{Flower13} around both low-mass (\citealt{Wampfler10, Benedettini12, Wampfler10, Karska13}) and high-mass (W11) YSOs. Even when its chemical origin is unclear, the presence of OH tends to be attributed to shock waves around low- and intermediate-mass YSOs (W13; \citealt{Dionatos13, Green13}). Observing OH is essential to quantifying the effects of the potentially intense radiation emanating from protostars and of the shock waves propagating in the surrounding regions. 

All previous studies of OH, except \citet{Wampfler10} and W11, were based on observations by the Kuiper Airborne Observatory or the \textit{ISO} satellite (as listed in W13), or by the PACS receiver onboard the \textit{Herschel} telescope and were at insufficiently high spectral resolution to enable the OH emission to be attributed unambiguously to shock waves. This paper presents the first velocity-resolved detection of three triplets of OH in the high-mass star-forming region Cep A with the GREAT spectrometer onboard SOFIA. We compare these observations with a grid of models, computed with the MHD shock code of \citet{Flower15}.

\section{Source selection}
\label{sec:sose}

Cepheus A (hereafter Cep A) is a well-known star-forming region, first observed by \citet{Sargent77}, \citet{Gyulbudaghian78} and \citet{Rodriguez80}. It is located in the Cepheus cloud, at a distance of about 700~pc \citep{Moscadelli09}. Cep A exhibits numerous manifestations of star formation activity, including peaks of CO emission, dense molecular clumps, H$_2$O and OH masers, hyper-compact \ion{H}{ii} regions, variable radio continuum sources, Herbig-Haro objects, H$_2$ emission, clusters of far-infrared sources, and Class I and II YSOs (see the extensive overview of the region provided in \citealt{Cunningham09}). The most spectacular feature is the gigantic anisotropic outflow structure, which is best seen in H$_2$ (e.g. \citealt{Hodapp94}), embracing several Herbig-Haro objects \citep{Cunningham09} and probably powered by the radio source HW2 \citep{Hughes82,Hughes84,Rodriguez94}, whose luminosity of 10$^4 L_\odot$ \citep{Garay96} implies a mass of 15--20~$M_\odot$. Additionnally, \citet{Cunningham09} have suggested that the submillimetre and radio continuum source HW3c might be driving the shocks of the HH168 (the brightest section of the outflow pointing towards the north-west) outflow component, and its counterflow. X-ray emission from the whole region has been mapped by the \textit{Chandra}/ACIS instrument, revealing three prominent sources in the Cep A east core and a number of other X-ray sources, located in or outside of this core, as appears to be typical of star-forming regions \citep{Pravdo09}. Figure~\ref{figure1} shows the complexity of the region, as seen by \textit{Spitzer}/IRAC and in H$_2$ at 2.12~$\mu$m.

\section{Observations}
\label{sec:obs}   

\begin{table*}
\caption{Characteristics of the OH transitions between the $^2\Pi_{1/2}$ $J = 3/2$ and $J = 1/2$ states, observed in emission, and of the OH transition between the $^2\Pi_{3/2}$ $J = 5/2$ and $J = 3/2$ states, observed in absorption. $A(B)\equiv A \times 10^B$. The `shift' column contains the velocity shift relative to the component with the largest Einstein $A$ coefficient. Source: JPL \citep{Pickett98}.}             
\label{table1}      
\centering                          
\begin{tabular}{l  c  c  c  c  c  c c}        
\hline           
\hline
triplet & transition & frequency & $A_{\rm{ul}}$ & $g_{\rm u}$ & $g_{\rm l}$ & $E_{\rm u}$ & shift   \\
properties & $F^{\prime}_{p^{\prime}} \rightarrow F_p$ & (GHz) & (s$^{-1}$) & & & (K) & (km s$^{-1}$)  \\
\hline
\hline
\tiny{signal} & $1+ \rightarrow 1-$ & 1837.7466 & 2.1(-2) & 3 & 3& 270.1 & 11.5   \\
\tiny{1838~GHz} & $2+ \rightarrow 1-$ & 1837.8168 & 6.4(-2) & 5 & 3 & 270.1 & 0.0  \\
\tiny{163.1~$\mu$m} & $1+ \rightarrow 0-$ & 1837.8370 & 4.3(-2) & 3 & 1 & 270.1 & -3.3   \\
\hline
\tiny{image} & $1- \rightarrow 1+$ & 1834.7355 & 2.1(-2) & 3 & 3& 269.8 & 1.9   \\
\tiny{1835~GHz} & $2- \rightarrow 1+$ & 1834.7474 & 6.4(-2) & 5 & 3 & 269.8 & 0.0    \\
\tiny{163.4~$\mu$m} & $1- \rightarrow 0+$ & 1834.7504 & 4.2(-2) & 3 & 1 & 269.8 & -0.5    \\
\hline
\hline
\tiny{signal} & $2- \rightarrow 2+$ & 2514.29873 & 1.4(-2) & 5 & 5 & 120.7 & 2.1   \\
\tiny{2514~GHz} & $3- \rightarrow 2+$ & 2514.31670 & 1.4(-1) & 7 & 5 & 120.7 & 0.0  \\
\tiny{119.2~$\mu$m} & $2- \rightarrow 1+$ & 2514.35349 & 1.2(-1) & 5 & 3 & 120.7 & -4.5    \\
\hline
\hline
\end{tabular}
\end{table*}

\begin{table*}
\caption{Observational parameters of the OH transitions between the $^2\Pi_{1/2}$ $J = 3/2$ and $J = 1/2$ states, observed in emission around 1835 and 1838~GHz, and of the OH transition between the $^2\Pi_{3/2}$ $J = 5/2$ and $J = 3/2$ states, observed in absorption around 2514~GHz.}             
\label{table2}      
\centering                          
\begin{tabular}{l  c  c  c  c  c  c c c c}        
\hline           
\hline
triplet & transition & frequency & beam size & observing & integration time & spectral resolution & beam & forward & $T_{\rm sys}$ \\
properties & $F^{\prime}_{p^{\prime}} \rightarrow F_p$ & (GHz) & ($''$) & date & (on source; s) & (km~s$^{-1}$) & efficiency & efficiency & (K) \\
\hline
\hline
\tiny{signal} & $1+ \rightarrow 1-$ & 1837.7466 & & & & & & &   \\
\tiny{1838~GHz} & $2+ \rightarrow 1-$ & 1837.8168 & 16.3 & 17/04/13 & 228 & 1.25 & 0.67 & 0.97 & 2689   \\
\tiny{163.1~$\mu$m} & $1+ \rightarrow 0-$ & 1837.8370 & & & & & & & \\
\hline
\tiny{image} & $1- \rightarrow 1+$ & 1834.7355 & & & & & & & \\
\tiny{1835~GHz} & $2- \rightarrow 1+$ & 1834.7474  & 16.3 & 17/04/13 & 228 & 1.25 & 0.67 & 0.97 & 2689 \\
\tiny{163.4~$\mu$m} & $1- \rightarrow 0+$ & 1834.7504 & & & & & & &  \\
\hline
\hline
\tiny{signal} & $2- \rightarrow 2+$ & 2514.298730 & & & & & & &  \\
\tiny{2514~GHz} & $3- \rightarrow 2+$ & 2514.316705  & 11.9 & 17/04/13 & 378 & 0.91 & 0.70 & 0.97 & 9687  \\
\tiny{119.2~$\mu$m} & $2- \rightarrow 1+$ & 2514.353490 & & & & & & &  \\
\hline
\hline
\end{tabular}
\end{table*}
   
The observations of the Cep A star-forming region were conducted with the GREAT\footnote{GREAT is a development by the MPI f\"ur Radioastronomie and the KOSMA$/$Universit\"at zu K\"oln, in cooperation with the MPI f\"ur Sonnensystemforschung and the DLR Institut f\"ur Planetenforschung.} spectrometer \citep{Heyminck12} during the SOFIA flight on 17 April 2013, as part of Cycle 1 of the programme. One position was observed with coordinates R.A.$_{[\rm{J}2000]}$=$22^h56^m17 \fs 9$, Dec$_{[\rm{J}2000]}$=$+62^\circ$01$'$49$\farcs$0, corresponding to the HW2 source. The receiver was tuned to the frequencies (1837.817~GHz and 2514.316~GHz) of the OH lines and of the CO (16--15) line (1841.346 GHz). The observed lines were the OH triplets around 1834.7, 1837.8, and 2514.3~GHz (see Tables~\ref{table1} and \ref{table2} for spectroscopic details; hereafter triplets at 1835, 1838, and 2514~GHz), and the CO (16--15) line (see Table~\ref{table3}). Because of spin-orbit interaction, the OH rotational levels are built within two ladders, $^2\Pi_{1/2}$ and $^2\Pi_{3/2}$. Each level is further split by $\Lambda$-doubling and hyperfine structure. The 1835 and 1838 GHz (163.4 and 163.1~$\mu$m) transitions are within the $^2\Pi_{1/2}$ ladder, whereas the ground state one at 2514~GHz (119.4~$\mu$m) is within the $^2\Pi_{3/2}$ ladder (see figure and references in \citealt{Wampfler10}). The receiver was connected to a digital XFFTS spectrometer \citep{Klein12}, providing a bandwidth of 2.5~GHz and resulting in the respective spectral resolutions for these lines: 1.25, 1.25, 0.91, and 1.24 km s$^{-1}$ (given in Tables~\ref{table2} and \ref{table3}). At these spectral resolutions, the rms uncertainties in the 1835, 1838 and 2514~GHz OH lines and for the CO (16--15) line were: 0.17, 0.12, 0.33, and 0.19~K, respectively. The observations were performed in double beam-switching mode with an amplitude of 80$''$ (or a throw of 160$''$) at the position angle of 135$^\circ$ and a phase time of 0.5 sec. The nominal focus position was updated regularly against temperature drifts of the telescope structure. The pointing was established with the optical guide cameras to an accuracy of $\sim$5$''$. The beam widths and efficiencies are indicated in Tables~\ref{table2} (for the OH transitions) and \ref{table3} (for the CO transitions). The data were calibrated with the KOSMA/GREAT calibrator \citep{Guan12}, removing residual telluric lines, and further processed with the CLASS software\footnote{http://www.iram.fr/IRAMFR/GILDAS}. This processing mostly consisted of linear baseline removal. For the OH 2514~GHz line, the continuum temperature was approximately $\sim$9~K, whereas all the other lines had a baseline at the $\sim$6~K level.

\begin{table}
\footnotesize
\caption{\footnotesize{CO (16--15) line and observational parameters. Spectroscopic data source:~CDMS, \citep{Mueller01,Mueller05}}}             
\label{table3}      
\centering                          
\begin{tabular}{c c}        
\hline
$A_{\rm{ul}}$ (s$^{-1}$) & 4.05$\times$10$^{-4}$ \\
$\nu$ (GHz)  & 1841.345506 \\
$E_{\rm u}$ (K) & 751.72 \\
$\lambda$ ($\mu$m) & 162.81 \\
\hline
beam size ($''$) & 16.3 \\
observing date  & 17/04/13 \\
integration time (on source; s) & 180 \\
spectral resolution (km~s$^{-1}$) & 1.24 \\
$B_{\rm eff}$ & 0.67 \\
$F_{\rm eff}$ & 0.97 \\
$T_{\rm sys}$ (K) & 2742 \\
\hline
\end{tabular}
\end{table}

\section{Results}
\label{sec:resu}

\subsection{OH in emission}
\label{sub:ohemi}

\begin{figure}
\centering
\includegraphics[width=0.49\textwidth]{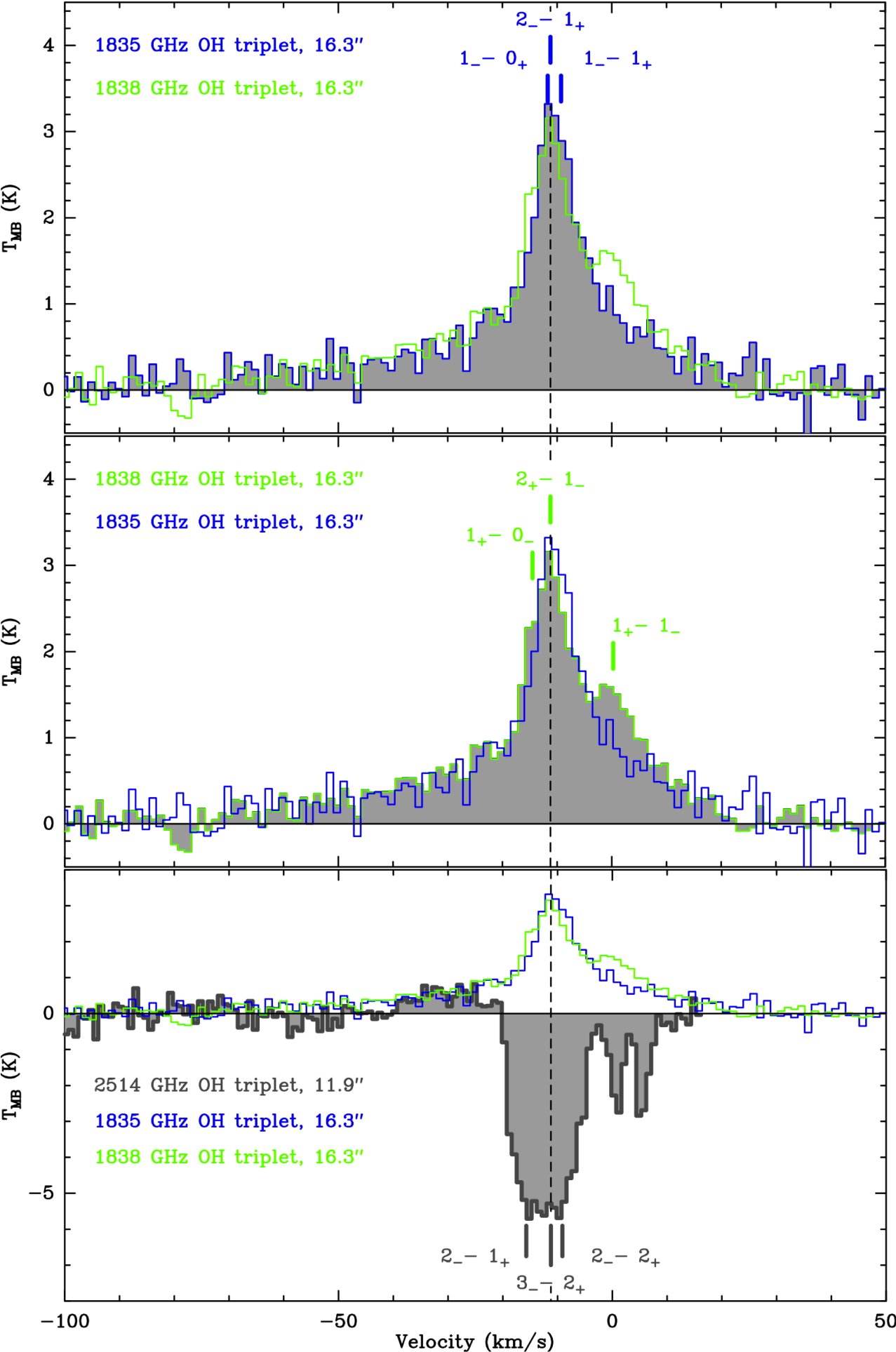}
\caption{\textit{Top panel:} 1835~GHz OH triplet (blue line and grey histogram), overlaid with the 1838~GHz triplet (green line). \textit{Middle panel:} The 1838~GHz OH triplet (green line and grey histogram), overlaid with the 1835~GHz triplet (blue line). \textit{Bottom panel:} The 2514~GHz OH triplet (dark grey and grey histograms), overlaid with the 1838~GHz (green line) and 1835~GHz (blue line) triplets. In all panels, the lines were observed at the Cep A HW2 position, indicated in Figure~\ref{figure1}, and the three components of the triplets are indicated in the same colour as the considered triplet: see Tables~\ref{table1} and \ref{table2}. The vertical dashed line is at the $\varv_{\rm lsr}$ of the cloud ($-11.2$~km~s$^{-1}$; e.g. \citealt{Narayanan96,Gomez99}). In the top and middle panels, the spatial and spectral resolutions are 16\farcs3 and 1.25~km~s$^{-1}$, respectively. In the bottom panel, they are 11\farcs9 and 0.91~km~s$^{-1}$.}
\label{figure2}
\end{figure}      

The emission in the two triplets at 1835 and 1838~GHz from the Cep A HW2 position can be seen in the two upper panels of Figure~\ref{figure2}. The frequencies of the components of each triplet are indicated; the 1$_+$--1$_-$ component of the 1838~GHz triplet is partially resolved. Both the triplets peak at the $\varv_{\rm lsr}$ of the cloud ($-11.2$~km~s$^{-1}$; e.g., \citealt{Narayanan96,Gomez99}), and exhibit high-velocity wings in the blueshifted and redshifted directions (respectively extending up to $\sim$--75 and to $\sim$35~km~s$^{-1}$), which are indicative of the presence of high-velocity shocks in the region. Since the 1838~GHz line profile is slightly more complex (the 1+$\rightarrow$1- component has a relatively large velocity shift of 11.5~km~s$^{-1}$, see Table~\ref{table1}), we applied a Gaussian fit to the 1835~GHz line alone, with no accounting for its hyperfine structure. We found that this line can be fitted by a combination of two Gaussian components: one narrow ($\Delta \varv \approx 8.4$~km~s$^{-1}$), peaking at --10.3~km~s$^{-1}$; the other broad ($\Delta \varv \approx 50.4$~km~s$^{-1}$), peaking at --11.8~km~s$^{-1}$. Such a double-Gaussian structure has been obtained by W13 in another massive SFR, W3-IRS5, whereas lower mass SFRs seem to exhibit only a broad component (e.g. \citealt{Wampfler10} for this conclusion, and  \citealt{Kristensen13} for the example of Ser SMM1). The quality of the fits of the OH lines is not decreased by requiring the peak velocities of all the Gaussian components to be --11.2~km~s$^{-1}$, the velocity of the source. The parameters of both of the double-Gaussian fits are given in Table~\ref{table4}.

\subsection{OH in absorption}
\label{sub:ohabs}

The profile of the triplet at 2514~GHz (see Table~\ref{table1}) from the Cep A HW2 position is shown in the bottom panel of Figure~\ref{figure2}. A baseline has been removed, since the continuum temperature at this frequency was $\sim$9~K (see Section~\ref{sec:obs}). The triplet structure is not resolved, but three velocity components can be seen in absorption. One peaks at the cloud velocity, showing the presence of OH in the ambient cloud; two other line-of-sight clouds must give rise to the absorption features at 1.2~km~s$^{-1}$ and 4.8~km~s$^{-1}$, although we have found no record in the literature of observations of the corresponding absorption components in other species. Neither of these velocity components is present in the other OH triplets. No evidence could be found of another species or line likely to cause absorption at these frequencies. Finally, the emission seen in the blue part of the 2514~GHz profile is at 2.5$\sigma$ level in several velocity channels between --35 and --25~km~s$^{-1}$. It coincides with the wings of the triplets seen in emission, despite the difference in angular resolution. It is therefore likely to be associated with the blueshifted shocked gas.

\subsection{CO (16--15) line emission}
\label{sub:co1615le}

\begin{figure}
\centering
\includegraphics[width=0.49\textwidth]{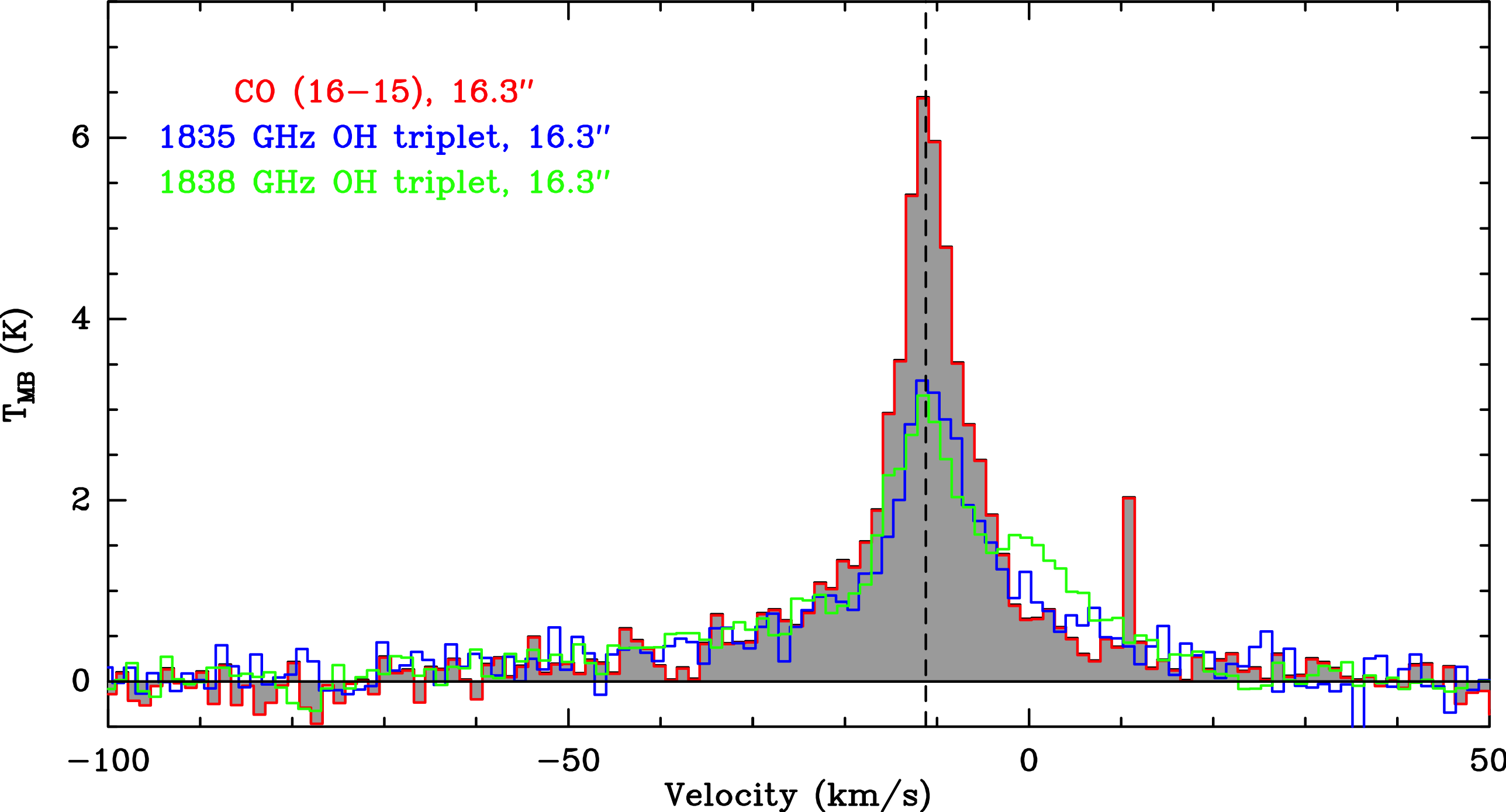}
\caption{CO (16--15) transition (red line and grey histograms) at the Cep A HW2 position, overlaid with the 1835~GHz and 1838~GHz triplets (blue and green lines, from Figure~\ref{figure2}). The spatial and spectral resolutions of the CO line are 16\farcs3 and 1.24~km~s$^{-1}$. The vertical dashed line is at the cloud $\varv_{\rm lsr}$ of $-11.2$~km~s$^{-1}$ \citep{Narayanan96,Gomez99}.}
\label{figure3}
\end{figure}

\begin{table*}
\caption{Parameters of the Gaussian decompositions of the OH 1835~GHz and CO (16--15) lines. `rms base' and `rms line'  denote the rms values associated with the residual. When fitting the lines with a double Gaussian, forcing the $\varv_{\rm peak}$ of one or both of the components to the $\varv_{\rm lsr}$ of the cloud leads to no more than $\sim$12\% variations in the rms values.}             
\label{table4}      
\centering                          
\begin{tabular}{l  c  c  c  c  c c c c c c c c c c }        
\hline
 & \multicolumn{5}{c}{double-Gaussian decomposition} &  & & \multicolumn{7}{c}{triple-Gaussian decomposition}  \\
 & \multicolumn{2}{c}{OH 1835 GHz} & &  \multicolumn{2}{c}{CO (16-15)} & & &  \multicolumn{3}{c}{OH 1835 GHz} & &  \multicolumn{3}{c}{CO (16-15)}  \\
 &  narrow & broad & &  narrow & broad & & & blue & ambient & red & &  blue & ambient & red \\
\hline
$\Delta\varv$ (km s$^{-1}$) & 8.4 & 50.4  & & 7.1 & 36.2 & &  & 53.6 & 7.8 & 15.2 & &  48.1 & 7.3 & 5.3 \\
$\varv_{\rm peak}$ (km s$^{-1}$) & -10.3 & -11.8  & & -10.9 & -12.7 & &  & -14.7 & -10.6 & -2.6 & &  -17.2 & -11.0 & -3.0 \\
$T_{\rm peak}$ (K) & 2.3 & 0.9  & & 4.9 & 1.2 & &  & 0.7 & 2.4 & 0.3 & &  0.8 & 5.3 & 0.7 \\
\hline
rms base (K) &  \multicolumn{2}{c}{0.17}   & & \multicolumn{2}{c}{0.19}  & &  &  \multicolumn{3}{c}{0.16}  & &  \multicolumn{3}{c}{0.19}  \\
rms signal (K) & \multicolumn{2}{c}{0.19}  & & \multicolumn{2}{c}{0.21}  & &  &  \multicolumn{3}{c}{0.18}  & &  \multicolumn{3}{c}{0.21}  \\
\hline
\end{tabular}
\end{table*}

We also obtained one velocity-resolved CO (16--15) spectrum in Cep A HW2 that we present in Figure~\ref{figure3}. The parameters of the line and of the telescope at this frequency are given in Table~\ref{table3}. The line has a similar structure to the OH triplets: it peaks at the $\varv_{\rm lsr}$ of the cloud and has line wings typical of shocks in the blueshifted and redshifted directions (respectively extending up to $\sim$--75 and to $\sim$35~km~s$^{-1}$). Very similar (double- or triple-) Gaussian decompositions can be applied to the OH line at 1835~GHz (see Table~\ref{table4}). This is after the narrow emission feature at $\sim$10.7~km~s$^{-1}$ is separately fitted by a single Gaussian component (with parameters $\Delta\varv \sim$1~km~s$^{-1}$, $\varv_{\rm peak}$ = 10.7~km~s$^{-1}$, and $T_{\rm peak} \sim$1.7~K). This feature is due to mesospheric CO over-compensated for by the correction for the atmospheric opacity. We found the CO line can be fitted by two Gaussians of width 7.1~km~s$^{-1}$ and 36.2~km~s$^{-1}$, centred at $-10.9$ and $-12.7$~km~s$^{-1}$, respectively (see Table~\ref{table4}). The CO line also peaks at the cloud velocity (contrary to the corresponding spectrum of Ser SMM1: \citealt{Kristensen13}). As for the OH 1835~GHz line, the quality of the fits of the CO line is not decreased by requiring the peak velocities of all the Gaussian components to be -11.2~km~s$^{-1}$, the velocity of the source. The similarity of the line wings in both of the OH emitting triplets and the CO lines suggests that the observed OH emission comes from the same gas as the CO (16--15) line and indicates the presence of high-velocity shocks in the observed region. The intensities integrated over the blue, red, and total velocity ranges (see Table~\ref{table1} for the corresponding values) are 43.5~K~km~s$^{-1}$, 41.9~K~km~s$^{-1}$, and 85.3~K~km~s$^{-1}$.

\section{Discussion}
\label{sec:disc}

\begin{table*}
\caption{Observed integrated intensities ($\int T_{\rm MB} \rm{d}\varv$, in K km s$^{-1}$) of the emission lines. The uncertainty in the integrated intensities mostly comes from the line temperature calibration of the order of $\pm 10$\%. LV and HV stand for low-velocity and high-velocity. The \lq total' value was used in scenario 1 (section~\ref{sub:disc1}), while the LV and HV values were used in scenario 2 (section~\ref{sub:disc2}).}  
\label{table5}      
\centering                          
\begin{tabular}{l  c  c  c  c  c  c}        
\hline           
\hline
 & \multicolumn{3}{c}{blueshifted} & \multicolumn{3}{c}{redshifted} \\
component & HV & LV & total & LV & HV & total \\
velocity interval (km s$^{-1}$) & [-90;-19] & [-19;-11] & [-90;-11] & [-11;-3] & [-3;30] & [-11;30] \\
corresponding $\Delta \varv$ (km s$^{-1}$) & 71 & 8 & 79 & 8 & 33 & 41 \\

\hline
\tiny{OH 1835~GHz} & 20.7 & 15.5 & 36.2 & 18.1 & 15.5 & 33.6   \\
\tiny{OH 1838~GHz} & 20.4 & 16.7 & 37.1 & 16.3 & 18.8 & 35.1   \\
\tiny{OH 2514~GHz} & 21.5\tablefootmark{a} & -- & -- & -- & -- & --   \\
\hline
\tiny{CO (16--15)} & 16.4 & 27.1 & 43.5 & 28.0 & 13.9 & 41.9 \\
\hline
\hline
\end{tabular}\\
\hspace{-6.8cm}\tablefoottext{a}{Integrated between -90 and -20~km~s$^{-1}$.}
\end{table*}

In this section, we present an analysis of our observations, based on a comparison of the integrated intensities of OH and CO lines with the values computed by the shock model of \citet{Flower13, Flower15}. The underlying assumption is that the OH and CO emission stems from shocked gas. We note that no H$_2$ emission is detected at 2.12~$\mu$m in our observed region. This does not contradict our assumption, but rather reflects the high extinction associated with the HW sources in the region ($A_{\rm v} = 500$ to 1000, \citealt{Cunningham09}). For instance, a microjet was recently imaged at high spatial resolution in the HH212 protostellar outflow by \citet{Codella14} in SiO and \citet{Podio15} in SO and SO$_2$, without having a counterpart in H$_2$ emission at 2.12~$\mu$m. Our modelling method is the same as  presented in \citet{Gusdorf11, Gusdorf12}. The model is one-dimensional, can simulate the propagation of stationary C- and J-type shocks, and includes a self-consistent Large Velocity Gradient ({\sc LVG}) treatment of the radiative transfer in the lines emitted by various cooling species (CO, H$_2$O, SiO, NH$_3$, and OH). The isotropic approximation was used for the escape probability with $\beta = (1 - e^{-\tau_\perp}) / \tau_\perp$, $\tau_\perp$ being the LVG opacity in the direction perpendicular to the shock front (e.g. \citealt{Gusdorf081}). Solving the radiative transfer in the lines within the shock model is more precise than an LVG post-processing of the shock model's output files. Indeed, it allows the level populations to be computed under the steady state assumption ($\rm{dn} / \rm{dt} = \varv \cdot \partial \rm{n} / \partial \rm{z}$) instead of under the statistical equilibrium one ($\rm{dn} / \rm{dt} = 0$). The range of input parameters covered by our calculations was as follows:  pre-shock densities $n_{\rm H} = n({\rm H}) + 2n({\rm H}_2) = 10^4$, 10$^5$, 10$^6$~cm$^{-3}$; transverse magnetic field strengths $B$($\mu$G) = $b[n_{\rm H}\rm{(cm^{-3})}]^{1/2}$, where the parameter $b = 1$ for C-type and $b = 0.1$ for J-type shocks; and shock velocities $\varv_{\rm s}$ = 10, 15, 20, 25, 30, 35~km~s$^{-1}$. In the high-density regions of the Cep A outflow, the relation between the total magnetic field strength and the density was observationally shown to be $B$($\mu$G)  $\propto n_{\rm H}\rm{(cm^{-3})}]^{0.47}$ by \citet{Vlemmings08}, based on a collection of OH, NH$_3$, CH$_3$OH, and H$_2$O maser measurements. We adopted the same law with a proportionality factor of 1 for the transverse magnetic field strength. Grain-grain interactions, such as studied by e.g. \citet{Guillet11}, are not included in our model. The maximum shock velocities in the grid are determined by the following considerations: C-type shock waves cannot propagate above a critical shock speed, which is, for example, 32~km~s$^{-1}$ for $b =$ 1 and  $n_{\rm H}$ = 10$^6$~cm$^{-3}$ \citep{Flower03}. Additionally, when H$_2$ -- the main coolant -- becomes dissociated, a thermal runaway occurs that prevents the model from converging towards a cold, compressed post-shock medium. In this case, no C-type shock can propagate, only a J-type one \citep{Lebourlot02}.%; at sufficiently high shock speeds -- about 50~km~s$^{-1}$ for J-type models -- a so-called `radiative precursor'  develops ahead of the shock wave, heating, dissociating and ultimately ionizing the pre-shock gas \citep{Hollenbach89} -- a phenomenon that is not included in the model employed here. %We are aware that grain-grain interactions that are neglected here can affect the structure of the shock layer for shock velocities above about  20--25~km~s$^{-1}$ \citep{Anderl13}. However, the consequences for the integrated intensities of OH lines are minor: see Figure B.4 of \citet{Anderl13}. 

\subsection{Scenario 1: one outflow, two shock layers}
\label{sub:disc1}

As in the case of other molecular species (e.g. SiO: \citealt{Gusdorf081, Anderl13}), the double-Gaussian decomposition of the OH line profiles (Section~\ref{sub:ohemi}) fails to account for the complexity of the observed emission. Most likely, the observed line profiles are produced by a collection of shocked layers. The statistical modelling of such a collection has been shown by \citet{Lesaffre13} to involve a number of free parameters. Given the limited number of observational constraints, we first adopted the simplifying assumption that the line profiles are the result of two shocks, propagating in the blueshifted and redshifted directions (as in \citealt{Gusdorf12}), together with an `ambient' component, and fitted the OH and CO line profiles by a combination of three Gaussians; the corresponding parameters are given in Table~\ref{table4}, and the corresponding figures are shown in the appendix (Figures~\ref{figurea1} and \ref{figurea2}). This approach resulted in slightly better residual rms than the double-Gaussian fits. This approach was used only to infer an estimate of the shock velocity: when a line is fitted by the combination of narrow and broad Gaussian components, it does not mean that the shock is only responsible for the broad emission. OH emission from the shock is expected at velocities close to the velocity of the source and contributes to the narrow component. 

The triple-Gaussian decomposition indicates that the OH spectral lines are slightly wider than the CO line. This could be due to the hyperfine structure of the OH lines or to the uncertainty on the fit and on the baseline correction. It also indicates a very high FWHM for the blueshifted Gaussian component: 53.6~km~s$^{-1}$ for OH and 48.1~km~s$^{-1}$ for CO, corresponding to a calculated full width at tenth maximum (FWTM) of 97.7~km~s$^{-1}$ and 87.7~km~s$^{-1}$, respectively. If generated by a single shock -- which is unlikely -- this magnitude of velocity is typical of shocks with radiative precursors, which lie outside the scope of our models. If generated by a collection of shocks, the number of observational constraints is too small to permit a meaningful analysis. We therefore decided to exclude this velocity component from our analysis in this first approach. On the other hand, the velocities observed in the redshifted component fall within the domain of applicability of the shock code: FWHM and FWTM of 15.2~km~s$^{-1}$ and 27.7~km~s$^{-1}$, respectively, for OH and 5.3~km~s$^{-1}$ and 9.7~km~s$^{-1}$ for CO. In the case of OH, the full width at one percent of maximum is 39.2~km~s$^{-1}$. Accordingly, we attributed a velocity between 27.7~km~s$^{-1}$ and 41~km~s$^{-1}$ to the red-wing shock; the latter value is slightly greater than 39.2~km~s$^{-1}$, reflecting the observed red-wing emission (from approximately $-11$ to $30$~km~s$^{-1}$, regardless of the type of Gaussian fit). The intensity was integrated over this velocity range and is given in Table~\ref{table1}. 

\begin{figure*}
\centering
\includegraphics[width=\textwidth]{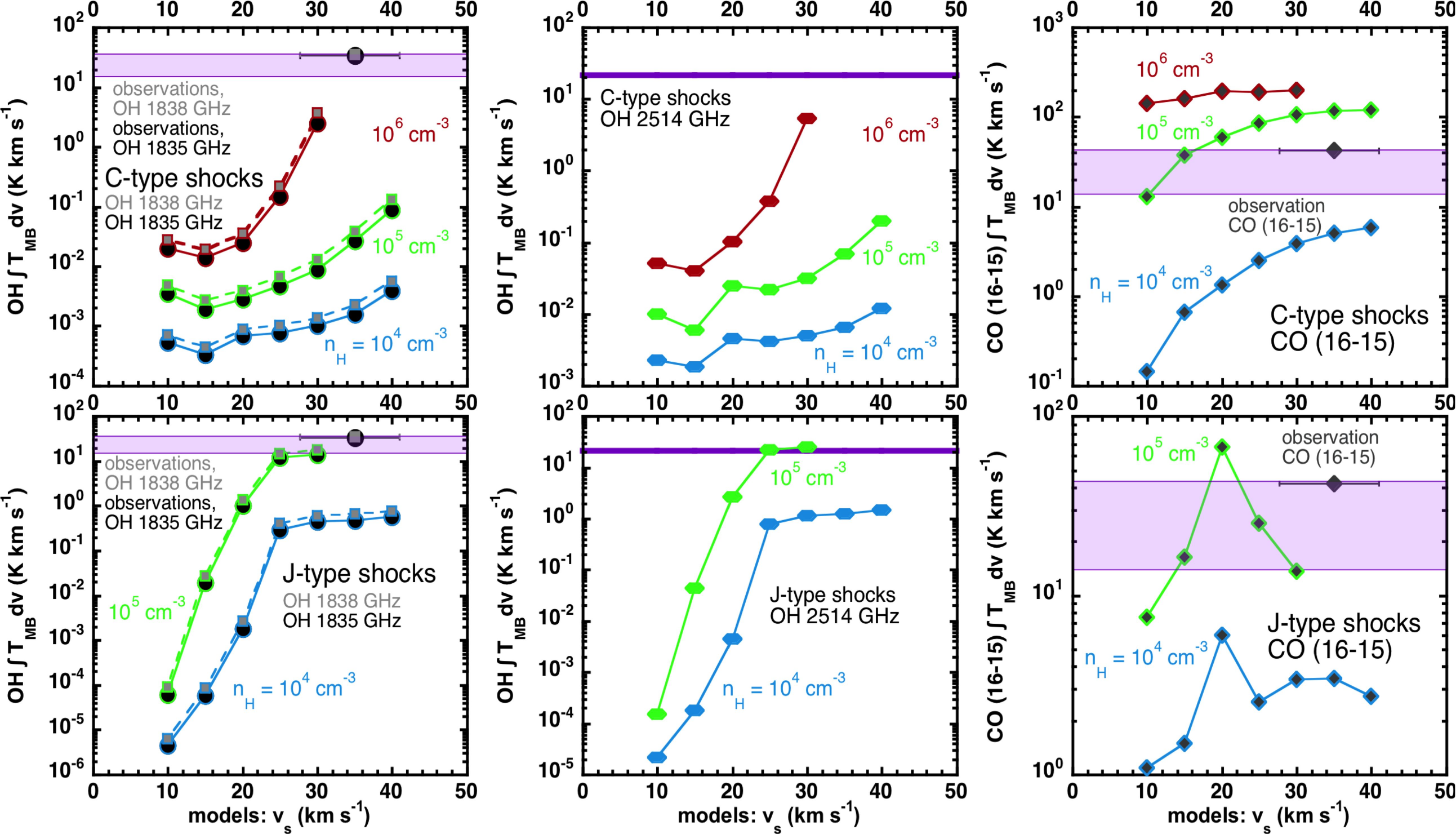}
\caption{Comparisons of the integrated intensities (Main beam temperature units) of all observed lines with the predictions of our grid of C-type (\textit{Upper panels}) and J-type (\textit{Lower panels}) models. In all panels, the abscissa is the shock velocity. In all panels, the colour symbols are the model predictions, where the red, blue, and green filled circles correspond to the pre-shock densities indicated in each panel. Finally, in all panels, the grey and black points are the observations integrated over the whole redshifted shock component (scenario 1, section~\ref{sub:disc1}), while the pink area corresponds to the range of values that correspond to the LV and HV values, see Table~\ref{table5}, in turn corresponding to scenario 2, section~\ref{sub:disc2}. In the central panels, the purple line is the observed intensity integrated between -90 and -20~km~s$^{-1}$. \textit{Left-hand side panels:} OH at $\sim$1835~GHz and $\sim$1838~GHz. The computed 1835~GHz and 1838~GHz intensities are connected by continuous and broken curves, respectively. The observed intensities are indicated by black (1835~GHz) and grey (1838~GHz) points with horizontal bars, reflecting the uncertainty on the shock velocity; the uncertainty on the intensity are of the order of the points sizes (see text). \textit{Central panels:} OH at $\sim$2514~GHz. \textit{Right-hand side panels:} CO (16--15).}
\label{figure4}
\end{figure*}

The comparison of the observations with the models is made in Figure~\ref{figure4}. In each panel, the abscissa is the shock velocity of the model. The (large) uncertainty in the observed shock velocity is indicated by the horizontal error bar. In the context of one-dimensional modelling, adopted here, the shock is assumed to be seen face-on. However, it is possible that shocks propagate in an already moving medium or that the flow is orientated towards the plane of the sky, resulting in, respectively, lower or higher shock velocities. In the former case, the observed integrated intensity should be compared to models with a lower velocity than the values inferred from our Gaussian fits, i.e. less than 27.7~km~s$^{-1}$. In the latter case, the velocity should exceed 41~km~s$^{-1}$ and is likely to lie outside the domain of applicability of the model, and hence of the scope of this study. Furthermore, both effects could intervene (an inclined shock propagating in a moving medium).

We note that varying the assumed inclination of the shock front has consequences for the line temperatures, calculated along the line of sight. In Appendix~\ref{sec:iiaai}, we provide a brief assessment of this effect for one J-type and one C-type model from the grid.

The ordinate in Figure~\ref{figure4} is the observed (or predicted) integrated intensity. The uncertainty in the integrated intensities is about the size of the filled circles that represent the observations. The upper and lower panels present C- and J-type shock models, respectively. The left-hand side, central, and right-hand side panels respectively present the comparisons for OH in emission (both at 1835 and 1838~GHz), OH in absorption, and CO (16--15). In the central panels, the purple line is the observed intensity, integrated between -90 and -20~km~s$^{-1}$. This velocity range corresponds to the blueshifted shock component and is not to be considered in this first part of our analysis. All the shock models in our grid predict somewhat more emission in the 1838~GHz than in the 1835~GHz line, as is observed. 

The main conclusion from Figure~\ref{figure4} is that, under the assumption that CO and OH arise from the same gas component, no shock model from our grid can fit all the observations, even with a filling factor equal to its upper limit of $1$. The trend that one perceives in this figure suggests that only a dense ($n_{\rm H} > 10^5$~cm$^{-3}$) and fast ($\varv_{\rm s} \gtrsim 30$~km~s$^{-1}$) J-type shock model could conceivably fit the data with a filling factor close to the unity for both OH and CO. Unfortunately we were unable to run such a case because of the numerical limitations of the model. These conclusions are not modified by varying the inclination angle, as indicated by our study presented in Appendix~\ref{sec:iiaai}. If confirmed, this would mean that the OH and high-$J$ CO emission originates in the strongest shock components within the beam, which are more likely to be in the jet than in the outflow cavity walls. We note that, in the densest and fastest J-type model ($n_{\rm H} = 10^5$~cm$^{-3}$ and $\varv_{\rm s} = 30$~km~s$^{-1}$), the optical depth of the OH line at 2514~GHz locally reaches high values ($\tau \approx$ 35--40). 

Under high-density, high-temperature conditions, one would expect grain-grain interactions (fragmentation, in particular) to play a significant part in the thermal structure of the shock (e.g. \citealt{Guillet09, Guillet11}). \citet{Anderl13} have shown that including them does not significantly alter the predicted OH emission of C-type models with $n_{\rm H}$ of the order of $10^5$~cm$^{-3}$ (see their Figure~B.4), but that it does affect the shape of the predicted energy distribution of the CO spectral line (see their Figure B.2). Furthermore, it is likely that, at such high shock velocities, a so-called radiative precursor develops ahead of the shock wave, which heats, dissociates, and ultimately ionizes the pre-shock gas \citep{Hollenbach89}; this phenomenon is not included in the model employed here. From the results in Figure~\ref{figure4}, we see that a dense and fast J-type solution could be combined with a less dense C-type solution (with $n_{\rm H}$ in the range [$10^4$--$10^5]$~cm$^{-3}$ and $\varv_{\rm s} \gtrsim 30$~km~s$^{-1}$), without significantly altering the comparison with the observations. This could mean that either the emission stems from a young, non-stationary shock or alternatively, that it stems from both the jet (J-type contribution) and the bowshock or the outflows cavity walls it has generated (C-type contribution). From this perspective, it is interesting to note that a solution based on a combination of C- and J-type models has been found by \citet{Flower032} to fit the molecular H$_2$ observations of \citet{Wright96} in a position (Cep A West) located 15$''$ away from the presently studied one. 

\subsection{Scenario 2: two outflows or one outflow with cavity walls, four shock layers}
\label{sub:disc2}

\citet{Cunningham09} discuss the possibility that two outflows could be present in the beam of their CO observations, one being responsible for low-velocity emission (less than 8~km~s$^{-1}$ away from the $\varv_{\rm lsr}$), the other for higher-velocity emission (more than 8~km~s$^{-1}$ away from the $\varv_{\rm lsr}$). They also indicate that this low-velocity emission could originate in the walls of a large-scale bipolar cavity. In either case, at least four shocks would be propagating within our observing beam, all with filling factors close to the unity. Lacking more precise information, we attribute the line emission to four velocity ranges -- one (so-called LV) of low- and one (so-called HV) of high-velocity -- for  the blueshifted and the redshifted emission. The intensity of the OH line at 2514~GHz, which is integrated between -90 and -20~km~s$^{-1}$ and indicated by a purple line in the central panels, thus roughly corresponds to the blue HV range. It could be a lower limit to the real value, owing to absorption at the velocities in the range closest to the source's velocity. The definitions of the velocity ranges and corresponding integrated intensities are given in Table~\ref{table5}.

Except for the 2514 GHz OH line, the line intensities, integrated over the different velocity ranges, are similar in magnitude. When presenting these results in Figure~\ref{figure4}, we show only the minimum and the maximum values of the integrated intensities in the four velocity ranges (the pink horizontal lines) and shade the interval thus defined in pink. We chose to extend the pink band over the whole shock-velocity range of the panels because the width of the HV or LV emission is not necessarily representative of the shock speed (owing to projection effects, or if the shock is propagating in already moving material).

From Figure~\ref{figure4} and Table~\ref{table5}, we draw the following conclusions: 
\begin{itemize}
\item For the HV blueshifted component, the wide velocity range (71~km~s$^{-1}$: see Table~\ref{table4}) implies that comparisons with our models are possible only if the shock is propagating in already moving material. In this case, the agreement with the observations is somewhat better than for scenario 1, Section~\ref{sub:disc1}, because the OH and CO integrated intensities are slightly lower, and because the blue wing of the 2514~GHz OH line can now be fitted by the models. A J-type shock with $n_{\rm H} = 10^5$~cm$^{-3}$, $b = 0.1$, and $\varv_{\rm s} = 30$~km~s$^{-1}$ is almost a good fit to the observations.
\item For the LV blueshifted and redshifted components, the integrated intensities in the various observed lines have similar values, and their velocity widths are similar. Comparisons with shock models show that, for an inclination angle in the range of 70--75$^\circ$, which transforms a speed of 8~km~s$^{-1}$ along the line of sight into a propagation speed of 25--30~km~s$^{-1}$, the observations can be reproduced, as may be seen in Figure~\ref{figure4}. In this case,  J-type shocks with $n_{\rm H} = 10^5$~cm$^{-3}$, $b = 0.1$, and $\varv_{\rm s} = 25-30$~km~s$^{-1}$ provide the best fit to the observations. However, the high value assumed for the inclination angle introduces additional uncertainty (see Figure~\ref{figureb1}).
\item For the HV redshifted component, a J-type shock with $n_{\rm H} = 10^5$~cm$^{-3}$, $b = 0.1$, and $\varv_{\rm s} = 30$~km~s$^{-1}$ fits the observed emission of both CO and OH.
\end{itemize}
In all these cases, the J-type solution could be combined with a less dense C-type solution (with $n_{\rm H}$ in the range [$10^4$--$10^5]$~cm$^{-3}$ and $\varv_{\rm s} \gtrsim 30$~km~s$^{-1}$) without significantly detracting from the quality of the fits to the observations, as was found in Section~\ref{sub:disc1}.

\section{Concluding remarks}
\label{sec:conc}

We have reported SOFIA observations of CO and OH spectral lines in the Cep A massive SFR. We considered two approaches in our analysis of the data, based on the CO study of \citet{Cunningham09}. In the first approach, we tried to fit one shock model per blueshifted and redshifted gas component. We found that no single model from the grid of \citet{Flower15} could adequately fit these measurements. This conclusion is at variance with the findings from studies of shocks associated with low-mass star formation. For instance, \citet{Karska142} have shown that C- and J-type models are generally capable of fitting the emission lines of various species, observed around numerous YSOs in the Perseus cloud. \citet{Leurini13, Leurini141} have shown that, in the massive SFR IRAS 17233--3606,  H$_2$O and SiO observations could be interpreted in terms of C-type shocks with a high pre-shock density ($n_{\rm H} \approx 10^6$~cm$^{-3}$) and shock velocity ($\varv_{\rm s} \approx 30$~km~s$^{-1}$). In the second approach, we assumed the existence of two outflows in the region or the coexistence of one outflow (associated with the high-velocity emission) with large-scale, bipolar cavity walls (associated to the low-velocity emission). We thus divided the lines into four velocity components and tried to fit shock models to each of them. We could fit the redshifted HV component successfully by a J-type shock with $n_{\rm H} = 10^5$~cm$^{-3}$, $b = 0.1$, and $\varv_{\rm s} = 30$~km~s$^{-1}$. This model could be combined with a less dense  C-type solution without significantly altering the comparison with the observations. This emission could arise in a young, non-stationary shock. Alternatively, it might originate in both the jet (J-type contribution) and either the bow shock or the outflow cavity walls (C-type contribution).

The present study confirms the necessity of recourse to high pre-shock densities when fitting molecular emission lines from massive SFR. A potentially significant limitation of our current models is that they exclude the effects of a UV radiation field, which might be emitted by the driving source or produced by the shock itself, and it is clear that progress will depend on further development of the shock models. We believe the observation of velocity-resolved OH spectra will prove to be a useful tool -- complementary to spectroscopic observations of H$_2$O and OI, e.g. at 63$\mu$m by SOFIA/GREAT -- when seeking to understand the water chemistry in high-mass SFRs.

\begin{acknowledgements}
      We thank an anonymous referee and Malcolm Walmsley for comments that helped to improve this paper. We thank the SOFIA operations and the GREAT instrument teams, whose support has been essential for the GREAT accomplishments, and the DSI telescope engineering team. Based [in part] on observations made with the NASA/DLR Stratospheric Observatory for Infrared Astronomy. SOFIA Science Mission Operations are conducted jointly by the Universities Space Research Association, Inc., under NASA contract NAS2-97001, and the Deutsches SOFIA Institut, under DLR contract 50 OK 0901. We thank N. Cunningham for providing us with the spectacular image of the region. This work was partly funded by grant ANR-09- BLAN-0231-01 from the French Agence Nationale de la Recherche as part of the SCHISM project. It was also partly supported by the CNRS programme \lq Physique et Chimie du Milieu Interstellaire'.    
      
\end{acknowledgements}

%\begin{thebibliography}{}

%  \bibitem[1966]{baker} Baker, N. 1966,
%\end{thebibliography}

\bibliographystyle{aa}
\bibliography{biblio}

%\Online
\begin{appendix} 
\section{Triple-Gaussian fits of the OH 1835 GHz and CO (16--15) lines}

\begin{figure}
\centering
\includegraphics[width=0.49\textwidth]{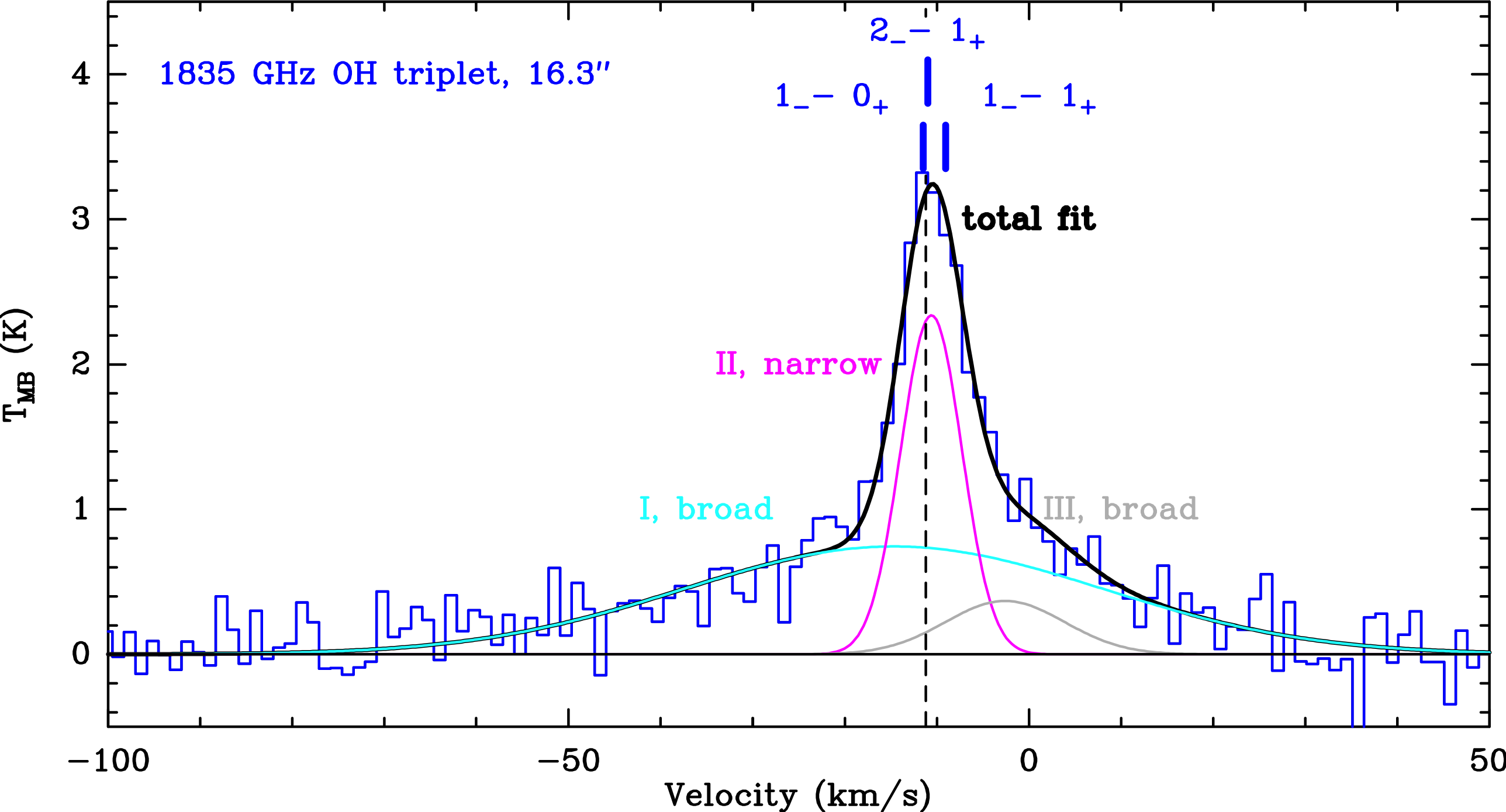}
\caption{Triple-Gaussian fit of the 1835~GHz OH triplet (blue line, from Figure~\ref{figure2}): blueshifted (light blue line), ambient (pink line), redshifted (grey line) components, and total fit (black line). The vertical dashed line is at the cloud $\varv_{\rm lsr}$ of $-11.2$~km~s$^{-1}$ \citep{Narayanan96,Gomez99}.}
\label{figurea1}
\end{figure}

\begin{figure}
\centering
\includegraphics[width=0.49\textwidth]{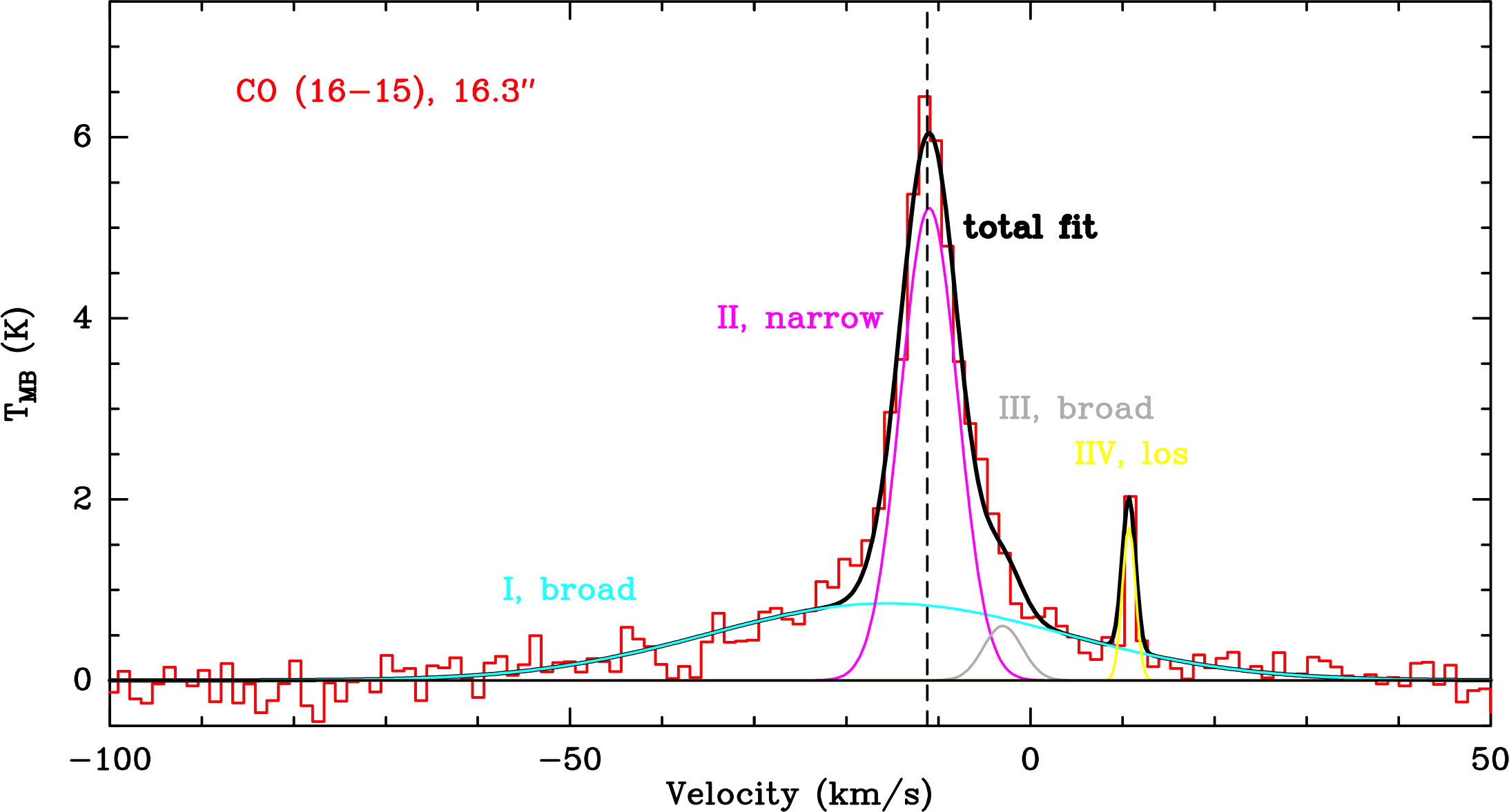}
\caption{Triple-Gaussian fit of the CO (16--15) transition (red line, from Figure~\ref{figure2}): blueshifted (light blue line), ambient (pink line), redshifted (grey line) components, and total fit (black line). An additional component has been fitted with respect to the OH line in Figure~\ref{figurea1}: a narrow CO feature at 10.7 km/s due to mesospheric CO over-compensated by the correction for the atmospheric opacity and well reproduced by a fourth Gaussian (yellow line). The vertical dashed line is at the cloud $\varv_{\rm lsr}$ of $-11.2$~km~s$^{-1}$ \citep{Narayanan96,Gomez99}.}
\label{figurea2}
\end{figure}

This appendix shows our Gaussian decomposition of the OH 1835 GHz and CO (16--15) emission lines in Figures~\ref{figurea1} and~\ref{figurea2}.

\section{Integrated intensities at arbitrary inclinations}
\label{sec:iiaai}

\begin{figure}
\centering
\includegraphics[width=0.49\textwidth]{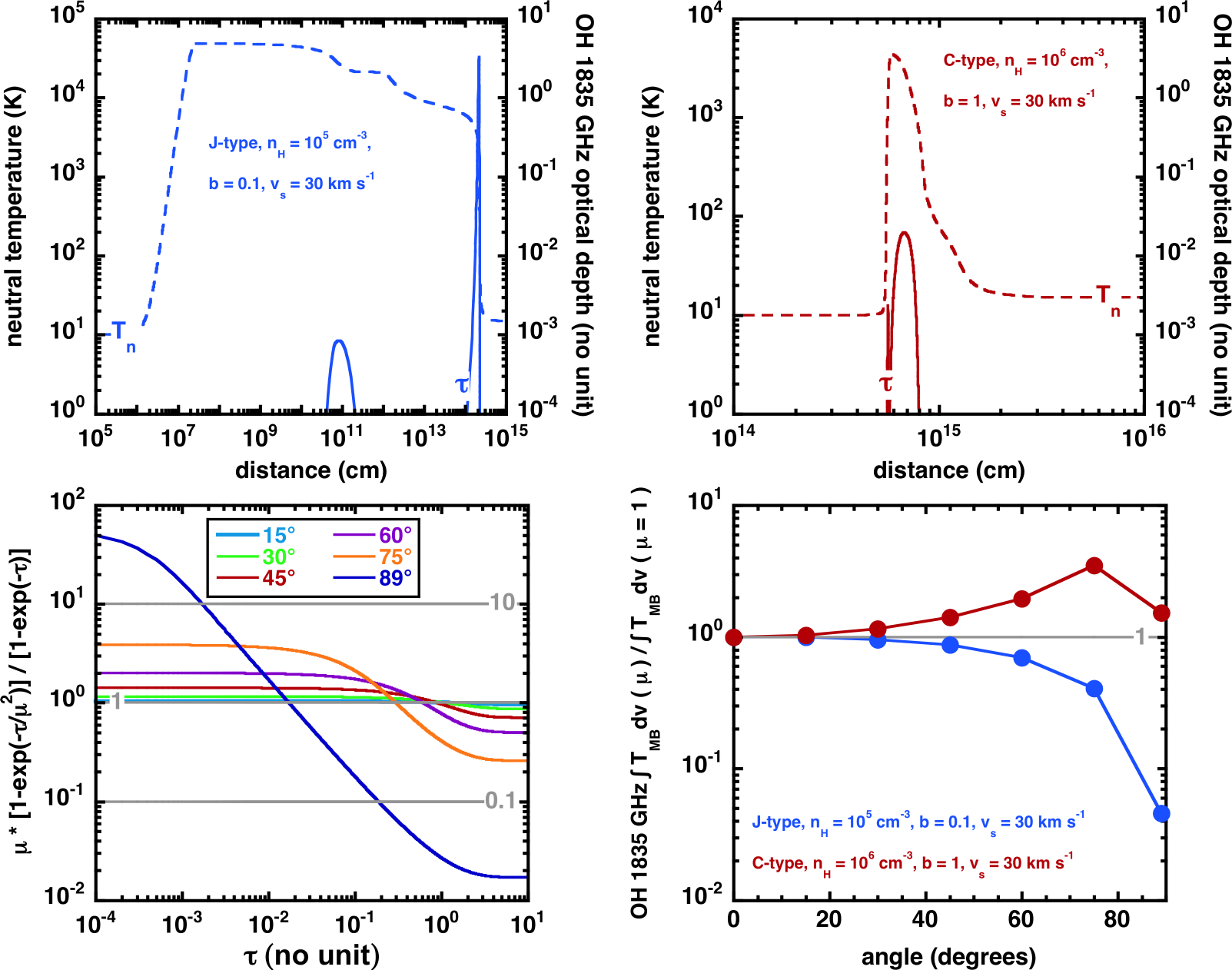}
\caption{\textit{Upper panels:} Evolution of the neutral temperature and optical depth of the OH line at 1835~GHz along the shock layer (versus a distance parameter along the direction of propagation of the shock) for a J-type model with $n_{\rm H}$ = 10$^5$~cm$^{-3}$, $b = 0.1$, and $\varv_{\rm s}=30$~km~s$^{-1}$ (\textit{left}) and a C-type model with $n_{\rm H}$ = 10$^6$~cm$^{-3}$, $b = 1$, and $\varv_{\rm s}=30$~km~s$^{-1}$ (\textit{right}). \textit{Lower left panel:} variation in the factor $\mu \cdot (1-{\rm e}^{-\tau/\mu^2})/(1-{\rm e}^{-\tau})$ with the optical depth for six inclination angles. \textit{Lower right panel:} variation in the ratio [$\int \rm{T} d\varv$ ($\mu$)]/[$\int \rm{T} d\varv$ ($\mu = 1$)] with the inclination angle for the two shock models considered.}
\label{figureb1}
\end{figure}

In this appendix, we briefly discuss the influence of inclination angle on the integrated intensity of the OH 1835 GHz line, thereby illustrating the difficulties encountered when comparing shock models with observations of OH emission lines in massive star-forming environments such as Cep A. We base our discussion on two particular shock models: a J-type model with $n_{\rm H}$ = 10$^5$~cm$^{-3}$, $b = 0.1$, and $\varv_{\rm s}=30$~km~s$^{-1}$ and a C-type model with $n_{\rm H}$ = 10$^6$~cm$^{-3}$, $b = 1$, and $\varv_{\rm s}=30$~km~s$^{-1}$. These models simulate the observations most closely (see Figure~\ref{figure4}). For these two cases, we show the variation in the neutral temperature and optical depth of the line considered through the shock layer (versus a distance parameter along the direction of propagation of the shock) in the upper panels of Figure~\ref{figureb1}. The line is optically thick in the region of the J-type shock, while it remains constantly optically thin in the C-type shock.

We consider an arbitrary inclination angle, $\theta$, and define $\mu = cos\,\theta$. Then, at each point in the shock layer, the line temperature is effectively multiplied by a factor ($1-{\rm e}^{-\tau/\mu^2}$)/($1-{\rm e}^{-\tau}$) (formula A.16 of \citealt{Gusdorf081}), where $\tau$ is the so-called LVG optical depth (see formula A.2 of \citealt{Gusdorf081}). Simultaneously, the velocity of the layer is multiplied by a projection factor along the photon path, $\mu$. In effect, the integrated intensity evaluated at each point of the shock should be multiplied by a factor $\mu \cdot (1-{\rm e}^{-\tau/\mu^2})/(1-{\rm e}^{-\tau})$. We display the variation in this factor with the optical depth for six inclination angles (15, 30, 45, 60, and 89$^{\circ})$ in the lower left-hand panel of Figure~\ref{figureb1}. This figure shows that the correction factor is less than 1 for optically thick transitions and greater than 1 for optically thin transitions, as might have been anticipated. Because the correction factor is applied point-by-point to the integrated intensity and depends on the inclination angle, we have not attempted to extend this study to all models of the grid.

The variation in the ratio [$\int \rm{T} d\varv$ ($\mu$)]/[$\int \rm{T} d\varv$ ($\mu = 1$)] with the inclination angle is shown in the lower right-hand panel of Figure~\ref{figureb1} for both shock models. In the J-type model, one can see that the  integrated intensity is reduced by the correction for the inclination angle, regardless of its value, because the line is optically thick in the post-shock region, where the emission is most significant. In the C-type model, the integrated intensity is increased by the correction for the inclination angle (with a maximum correction factor of $\sim$3.5 for an angle of 75$^{\circ}$), because the line is always optically thin. However, this correction is insufficient to fully reconcile the integrated intensity with the observed value in the framework of the discussion in Sections~\ref{sub:disc1} or~\ref{sub:disc2}.

\end{appendix}

\end{document}